\documentclass[conference]{IEEEtran}
\pdfoutput=1
\usepackage{cite}
\usepackage[keeplastbox]{flushend}
\usepackage{amsmath,amssymb,amsfonts}
\usepackage{algorithmic}
\usepackage{graphicx}
\usepackage{textcomp}
\usepackage[dvipsnames]{xcolor}
\usepackage{fancyhdr}
\usepackage[hyphens]{url}
\usepackage{pifont}
\usepackage{booktabs}
\usepackage{multirow}
\usepackage{xspace}
\usepackage[export]{adjustbox}
\usepackage[caption=false]{subfig}
\usepackage[pdfusetitle, pdfauthor={Chang Hyun Park, Ilias Vougioukas, Andreas Sandberg, David Black-Schaffer}]{hyperref}

\def\BibTeX{{\rm B\kern-.05em{\sc i\kern-.025em b}\kern-.08em
    T\kern-.1667em\lower.7ex\hbox{E}\kern-.125emX}}

\makeatletter
\newcommand*\etal[0]{{\it et~al}\@ifnextchar{.}{}{.\@\xspace}}
\makeatother

\newcommand\kB[0]{\,KB\xspace}
\newcommand\MB[0]{\,MB\xspace}
\newcommand\GB[0]{\,GB\xspace}

\hypersetup{
  pdfborder={0 0 0}, 
}

\IEEEoverridecommandlockouts %
\IEEEaftertitletext{\vspace{-1.5\baselineskip}}

\fancypagestyle{empty}
{
  
   \fancyhf{}
   \fancyfoot[C]{{This paper is distributed under the \href{https://creativecommons.org/licenses/by-nc-nd/4.0/}{CC BY-NC-ND} license}}
}

\begin{document}

\title{Page Tables: Keeping them Flat and Hot (Cached)}

\author{ \IEEEauthorblockN{Chang Hyun Park}
\IEEEauthorblockA{Uppsala University\\
chang.hyun.park@it.uu.se}
\and
\IEEEauthorblockN{Ilias Vougioukas}
\IEEEauthorblockA{Arm Research\\
ilias.vougioukas@arm.com}
\and
\IEEEauthorblockN{Andreas Sandberg}
\IEEEauthorblockA{Arm Research\\
andreas.sandberg@arm.com}
\and
\IEEEauthorblockN{David Black-Schaffer}
\IEEEauthorblockA{Uppsala University\\
david.black-schaffer@it.uu.se}
}

\maketitle
\thispagestyle{empty}
\pagestyle{plain}


\begin{abstract}

As memory capacity has outstripped TLB coverage, large data applications suffer from frequent page table walks. 
We investigate two complementary techniques for addressing this cost: reducing the number of accesses required and reducing the latency of each access. The first approach is accomplished by opportunistically ``flattening'' the page table: merging two levels of traditional 4\kB page table nodes into a single 2\MB node, thereby reducing the table's depth and the number of indirections required to search it. The second is accomplished by biasing the cache replacement algorithm to keep page table entries during periods of high TLB miss rates, as these periods also see high data miss rates and are therefore more likely to benefit from having the smaller page table in the cache than to suffer from increased data cache misses.

We evaluate these approaches for both native and virtualized systems and across a range of realistic
memory fragmentation scenarios, describe the limited changes needed in our kernel implementation and
hardware design, identify and address challenges related to self-referencing page tables and kernel memory allocation, and compare results across server and mobile systems using both academic and industrial simulators for robustness.

We find that flattening does reduce the number of accesses required on a page walk
(to 1.0), but its performance impact (+2.3\%) is small due to Page Walker Caches (already 1.5 accesses). Prioritizing caching has a larger effect (+6.8\%), and the combination improves performance by +9.2\%.
Flattening is more effective on virtualized systems (4.4 to 2.8 accesses, +7.1\% performance),
due to 2D page walks.
By combining the two techniques we demonstrate a state-of-the-art +14.0\% performance gain and -8.7\% dynamic cache energy and -4.7\%
dynamic DRAM energy for virtualized execution with very simple hardware and software changes.

\end{abstract}


\section{Introduction}
\textbf{The problem: Traditional page walks do not scale well with large data sets.}
While memory capacity has grown by $100\times$ 
over the past decade, TLB sizes have merely tripled to around 1500 L2 TLB entries, delivering a reach of only 3\GB with 2\MB large pages.
As a result, applications that use the large amount of available physical memory often suffer from significant numbers of TLB misses and resulting page walk delays.
Virtualized environments see an even larger penalty for the 2D page walk to translate each level of the guest page walk on the hypervisor side~\cite{2d_translation}.
This problem will be exacerbated with 5-level page tables for larger memories~\cite{intel:5_level}.

\begin{figure}[t]
\includegraphics[width=\columnwidth]{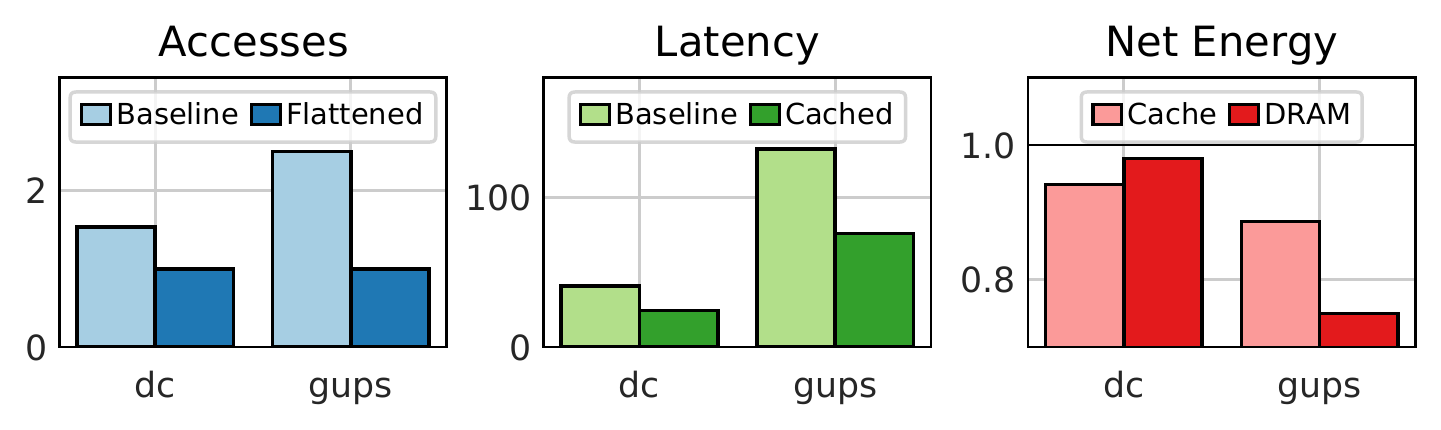}
	\caption{Left: reduction in memory requests per page walk from \textit{flattening the page table}.
	 Center: reduction in page walk latency from \textit{prioritizing
	caching page table entries}.
	Right: reduction in dynamic energy of the cache hierarchy and DRAM from flattening and
	prioritizing.
	For benchmarks with high/low (gups/dc) TLB miss rates
}
\label{fig:motiv_num_walks}
\end{figure}

\textbf{Background: Traditional page table trees have significant overheads on today's systems.}
\label{sec:intro:not_great}
Traditional page tables were designed with the assumption that memory is managed in contiguous blocks of exactly one page.
The practical implication of this is that nodes in the page table are the same size as a page, leading to a multi-level deep tree of 4\kB page table nodes, each mapped to its own 4\kB memory page.
Having each page table node be one memory page vastly simplifies allocation as the operating system does not have to maintain a separate reserve of larger blocks to guarantee it can allocate larger page table nodes.
Keeping the nodes small also avoids wasting memory due to fragmentation.
However, the small size of the page table node allocations combined with today's large memory capacities results in a deep page table tree, which incurs multiple serial memory indirections on each page table walk.

\textbf{Solution: Flatten the page table to reduce indirections and prioritize caching the page table to reduce latency.}
While there have been many proposals to increase the effective TLB coverage to avoid page walks, we instead seek to reduce the cost of page walks. 
First, we reduce the number of architectural memory accesses in a page table walk (typically 4 without virtualization or 24 with virtualization enabled).
This is done by \emph{flattening the page table} to reduce the number of indirections required for a
page walk by using large pages to store more entries in each page table node.
Second, we reduce the latency of the accesses by increasing the effectiveness of on-chip caches for the page table.
This is achieved by \emph{preferentially caching page table entries} during periods of high TLB
misses and low data reuse.
Together, we reduce the number of memory accesses per translation to 1
(non-virtualized) or 3 (virtualized), and reduce average walk latencies from 50.9 cycles down to
29.1 cycles.

Flattening the page table is based on two observations. 
First, the radix nature of the page table tree structure allows two levels of standard (4\kB) page table nodes to be combined into one level of large (2\MB) nodes.
This reduces the tree's depth and the number of indirections required to walk the tree, while leveraging existing 2\MB page support.
And, second, the page table itself is so much smaller and more static than the actual data, that the bloating~\cite{perforated_pages,ingens} and variable latency~\cite{ingens,illuminator} problems that plague 2\MB pages for data are not significant. 
To take advantage of these observations, we flatten the page table using large pages for nodes. 
To ensure that the OS is able to reliably allocate page table nodes, we make flattening optional on a per page table node basis.
Through a prototype, we find that the OS changes required to implement this new arrangement are small.

Concretely, for an 8\GB application we can reduce the number of pages in the $\approx$16\MB page table from  4106 4\kB pages with a standard 4-level table to only nine 2\MB pages in a flattened 2-level table, thereby reducing the number of  indirections required on TLB misses from 4 to 2. 
While this incurs a  $\approx$2\MB overhead ($9\times2$\MB vs. $4106\times4$\kB), the cost is negligible for large applications because the total size of the page table is so small (18\MB vs. 8\GB data).

Prioritizing caching page table entries is also based on two observations.
First, the page table itself is close to the size of the LLC, and, second, high TLB miss rates are correlated with high data miss rates.
This implies that we can expect to keep most of the page table in the cache and that doing so will not significantly hurt data access latency.
We find that preferentially keeping page table entries in the cache during phases of high TLB miss rates is a simple and effective way to reduce the latency of page walk accesses, and, when combined with flattening, we can achieve single-access cache hits for most page walks.

\textbf{Context: Page Walker Caches are excellent.}
Page walk caches (PWCs) already reduce the theoretical 4 memory system accesses per
page walk to $<1.5$ on average (max 2.5 on our random access benchmark), and from 24 to 4.4 for virtualized systems.
Flattening reduces this to 1 (2.8 virtualized), while cache prioritization reduces the latency of each access.

\textbf{Contributions.}
The impact of our approach for applications with relatively few TLB misses (dc) and many (gups) is shown in \autoref{fig:motiv_num_walks}. 
While flattening significantly reduces the number of memory requests per page walk (left), by itself
it has limited  performance benefit. However, for virtualized systems, and with realistic memory
fragmentation, its impact increases significantly (\autoref{sec:virt}), as they have more complex page walks and more pages in the their page tables.
Prioritization significantly reduces latency (middle) by avoiding most DRAM accesses for page walks
and does so without significantly hurting the overall cache performance (\autoref{sec:prio}).
When combined, the approaches beat the state of the art for performance and deliver significant reductions in dynamic cache and DRAM energy (right).
Our contributions:

\begin{itemize}
\item We identify, evaluate, and combine complementary approaches for reducing the impact of page walks: flattening to reduce the number of accesses and cache prioritization to reduce their latency.
\item We identify and quantify the importance of handling large allocation failures in the kernel on real systems.
\item We demonstrate the flexibility to dynamically choose where in the page table to flatten to efficiently support large data pages and evaluate its impact across three fragmentation scenarios.
\item We quantify the benefits for virtualized systems and explore the trade-offs in flattening the page table for the host, guest, or both in virtualized systems.
\item We show that flattened tables are not naturally compatible with recursive page tables and provide an efficient dereferencing solution.
\item We demonstrate the limited OS changes required by reporting on the code changes for a Linux implementation of a flattened page table.
\item We present simulations from server and mobile system on academic and industrial simulators for robustness.
\end{itemize}


\section{Related Work}
There has been a tremendous amount of work aimed at improving translation
range and efficiency (and thereby reducing the number of page
walks)~\cite{tlb_survey,colt,basu,clusterTLB,RMM,perforated_pages,tailored,enigma,delayed_trans,HTC,
ca_spot,multi_page_sizes,RMM,non_contig_superpages,v_ds, dead_tlb_cache}.
Other works have focused on reducing the TLB miss penalty by
improving the page table walk
caches~\cite{skipdontwalk,large_reach,2d_translation}, using
speculation to hide latency~\cite{spectlb, ahn_flat_nested,both_ways,ca_spot},
optimizing hash page tables~\cite{hash_cache},
and replicating page tables across NUMA nodes~\cite{mitosis}.
For virtualized systems, Gandhi~\etal proposed merging the 2D page table into a single dimension where possible~\cite{agile_paging}.
Ahn~\etal~proposed flat host page tables for precisely the virtual machine memory size~\cite{ahn_flat_nested}, which will perform similarly to our host page table flattening without the benefits of our guest-flattening.
Ausavarungnirun~\etal~proposed bypassing the shared cache for the lower levels of the tree to avoid pollution in throughput-oriented GPU systems~\cite{mask}.
Mazumdar~\etal~proposed predicting dead TLB entries and dead page table entries in the LLC~\cite{dead_tlb_cache}.
This could be used to extend our cache prioritization, although we have not investigated this.
Below we discuss four works that seek to directly reduce the cost of page walks.

Ryoo~\etal proposed POM\_TLB~\cite{pom_tlb} to use a part of the DRAM as a large
set-associative TLB.
This requires only a single memory lookup, and the entries can be cached.
On a miss in the in-DRAM TLB, a conventional page table walk is required.
Because this space is allocated at system boot, the required large contiguous partition can be guaranteed.
This approach has the benefit of not requiring any OS changes, but comes at the
cost of the complexity of scanning the structure at address space teardown, and
possible interference or security implications as the on-DRAM TLB is shared by all cores,
processes, and VMs.
Marathe~\etal extended POM\_TLB with a cache prioritization extension (CSALT~\cite{csalt}), to keep page table entries in the cache during frequent context switches.

Margaritov~\etal proposed ASAP~\cite{asap}, which prefetches the lower levels of the page table during the time spent accessign the higher levels.
However, as modern PWCs are effectively eliminate the accesses to the higher levels, our simulations show  very little opportunity for such prefetching.
ASAP requires that appropriate pages are stored sequentially in memory to enable prefetching without pointer chasing.
This requires that the kernel can allocate contiguous segments of memory for page table entries for prefetching to function, which is difficult to guarantee.
If such regions are not available, ASAP is unable to prefetch.

Elastic Cuckoo Hashing~\cite{ECH} restructures the page table into a hash table to generate lookup addresses without pointer chasing.
ECH uses multiple hashed regions for parallel lookups, which requires the OS to allocate large contiguous blocks upon creating or resizing the page table.
Since the OS cannot guarantee large contiguous memory allocations, this would make it difficult to implement in practice.

Compendia~\cite{compendia} is a parallel work that addressed flattening the page table.
They claim roughly twice the performance benefits that we observed. However, their methodology uses
saved page walk cycle estimates to compute performance change rather than simulation, and does not
include overlapping page walks, a data cache hierarchy, or realistic memory
fragmentation~\cite{quicksilver,perforated_pages}, which may account for the differences. Further,
they do not claim a kernel prototype, address self-referencing page tables, or compare to previous proposals.

Flattening alone achieves comparable main memory access reductions to what ECH could achieve (with its way-prediction), but without the complexity of dynamically resizing the hash table, is better than what POM\_TLB could achieve, as its cache does not cover the full page table, and is simpler than ASAP, as the layout modification leverages existing large page support.
Importantly, flattening provides a graceful fallback to 4\kB pages when larger contiguous
allocations fail, which our prototype kernel shows happens on heavily-loaded systems (\autoref{sec:impl:sw}). As a result, proposals that require large contiguous allocations to function (such as ECH) face sever implementation challenges.
Our final proposal to combine flattening and cache prioritization goes beyond CSALT's prioritization as it does not require a separate POM\_TLB cache to achieve similar results, and improves on Compendia in both energy and performance by coordinating with the memory hierarchy.


\section{Flattening the Page Table}

Page tables are organized as trees that entail a series of memory indirections for each page walk.
These pointer-chasing accesses lead to long latencies to satisfy TLB misses.
We can reduce the number of levels in the tree (flatten it) to reduce the number of indirections by using larger nodes in the tree.
When combined with modern page walker caches (PWCs), this achieves effective page walks of a single
memory access. In \autoref{sec:prio} we will add preferential caching to make  this single access a cache hit.

\subsection{Conventional page tables}
\begin{figure}[t]
\includegraphics[width=\columnwidth]{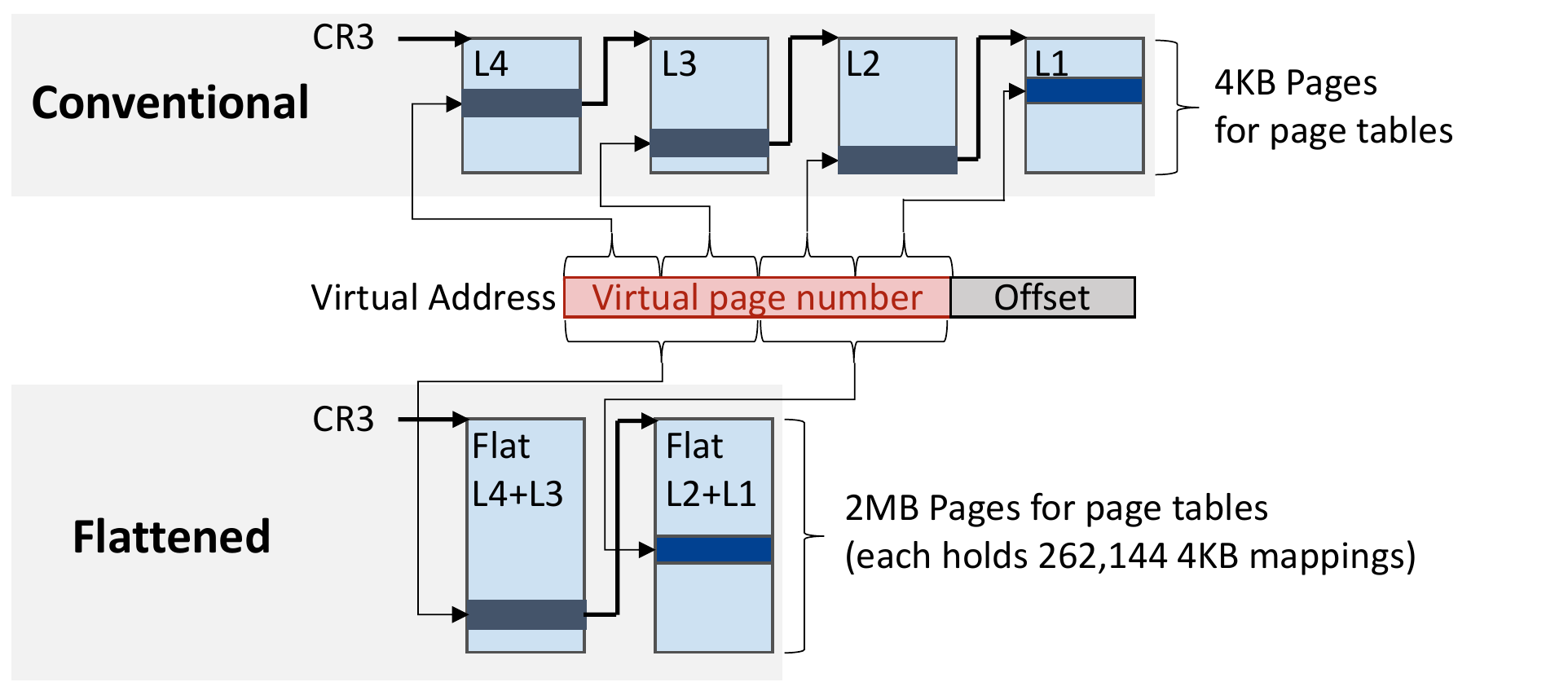}
	\caption{Conventional 4-level page table (top) and our proposed 2-level flattened page table (bottom). We utilize 2\MB pages for the flattened L4+L3 and flattened L2+L1 page table levels, however flattening can be applied using many combinations of natively supported page sizes.}
\label{fig:flattened_pt}
\end{figure}

We consider the conventional 64-bit x86 and Arm page tables, which use a page size of 4\kB and entry size of 8\,B, leading to a 512-ary radix tree. However flattening can be applied to other configurations.
\autoref{fig:flattened_pt} illustrates address translation pointer chasing through the page table tree for conventional (4\kB nodes, top) and one flattened (2\MB nodes, bottom) page table configuration.
The conventional tree has 512 entries per level, with each 9-bit segment of the address used to index within each level\footnotemark, and the address of the first page table node in the CR3 (x86) or TTBR (Arm) register.

\footnotetext{We label the page table L4, L3, L2 and L1 from root
to leaf.}

While an L1 (leaf) page table entry represents a 4\kB virtual memory region, each entry in higher-level nodes can either point to a lower-level node or directly represent a translation of a larger region: e.g., L2 entries can either point to an L1 node of 512 4\kB translations or directly translate a contiguous $512\times 4$\kB = 2\MB region.
Thus the page table elegantly supports large pages by tracking whether each entry is a pointer to a
lower-level node or direct translation. Some TLB designs leverage this fact to store
partial walks and final translations in the same HW structure (e.g. L2 TLB)~\cite{a76_trm, amd_opt}.

\subsection{Flattening the page table}
\begin{figure*}[t]
\centering
\includegraphics[width=1.0\textwidth]{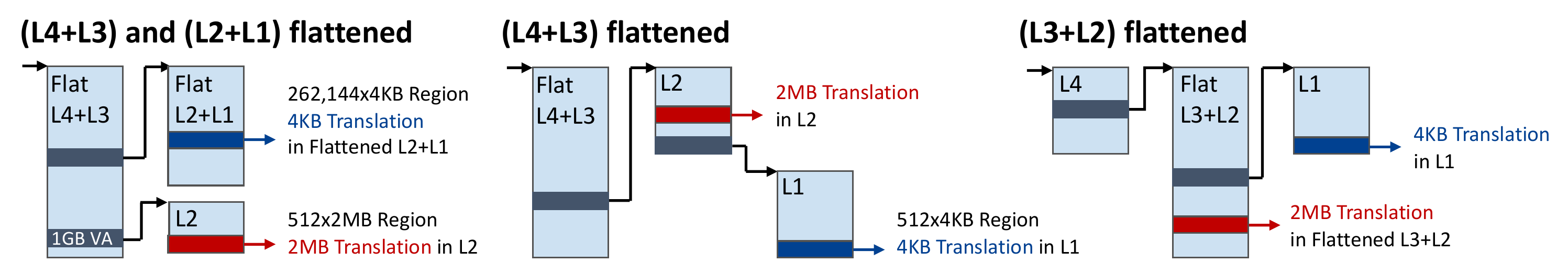}
	\caption{Flattening the L4+L3 and L2+L1 levels (left): 4\kB pages (flattened L2+L1) require only two accesses while 2\MB pages are mapped to 1GB VA ranges that do not flatten their L2+L1 (unflattened L2).
	Other approaches may flatten the first two levels (middle, L4+L3) or the middle two levels (right, L3+L2).
	 Our prototype OS implementation targets flattening L3+L2 (right) while our evaluations look at flattening both L4+L3 and L2+L1 (left). }
\label{fig:large_page_flattened}
\end{figure*}

Flattening uses the radix nature of the page table to naturally merge levels of the page table into single, larger levels, resulting in a shallower tree.
Merging two levels results in page table levels that naturally use the next larger page size.
For example, consider merging the L2 and L1 levels in the page table.
As each 4\kB L2 node points to 512 4\kB L1 nodes, and each L1 node provides 512 translations, if we flatten the L2 and L1 levels, we would replace each L2 node and its 512 L1 child nodes with a single 2\MB flattened node, which covers all $512\times512$ (262\,144) translations (\autoref{fig:flattened_pt}, bottom).
If all translations in the region are mapped, this approach saves the 4\kB of space needed to hold the original L2 node.
However, if some L1-sized regions of the translations are not mapped, this approach will waste 4\kB of space for each such region.
In the traditional page table, L1-sized regions that are not used would not have allocated 4\kB page table nodes, thereby saving memory.
This bloating in the page table itself is a side-effect of flattening the page
table, but as the page table itself is a minuscule fraction of the size of the
actual data (roughly 1/512th the size with regular pages and a contiguous VA space), the overhead is
negligible.

The choice of which page table levels to flatten is flexible.
In this work, we use the design shown in the bottom of \autoref{fig:flattened_pt} where we flatten the first two (L4+L3) and
the latter two (L2+L1) levels into 2\MB page table nodes.
However, one could flatten L3+L2 instead\footnotemark, which would be beneficial in the presence of 2\MB
data pages (shown in \autoref{fig:large_page_flattened}, right), or possibly even flatten the top
three levels (L4+L3+L2) using a 1\GB page table node that points to 2\MB data pages directly.
Alternatively, one could flatten the top two levels (L4+L3) using 2\MB page table nodes and allocate
all memory in 1\GB data pages.  These approaches all have different trade-offs in terms of bloating
within the page table itself and the ease of creating sufficiently large allocations.  The choice of
which levels to flatten can be made at on a per-process, and per-table, basis at runtime by the OS with or without application hints.
\footnotetext{We have also simulated this flattening on both our simulators but omit them from the
paper due to space constraints. Representative results can be seen in the mobile case study in \autoref{fig:mobA}.}

Because they are based on the radix nature of the page table, flattened page tables provide a key characteristic for practical implementation: graceful fallback to conventional page tables when needed.
If the OS is unable to allocate a 2\MB page for a flattened page table node, it can instead allocate the standard two levels of 4\kB page table nodes at any place in the page table with no additional overhead.
This allows the OS to choose whether to take the time at allocation for compaction,
merge the two levels at a later point when a 2\MB page is available, or simply use the standard two-level approach.

Only small changes are needed to support flattened page tables.
The system architecture needs to be able to inform the hardware page walker of which nodes are flattened
with one bit in one of the control registers for the root and one bit in each page table entry (two bits to support 1\GB flattened nodes).
The hardware page table walker can then read those bits to determine which parts of the virtual address should be used for index bits at each level. (See \autoref{sec:impl:hw}.)

\subsection{Page Walk Caches}
\label{sec:flatten:psc}
Page walks are cached in three ways: translation caches (TLBs), page table entries in the regular
data cache hierarchy, and through Page Walker Caches (PWCs), such as Intel's Paging Structure
Cache~\cite{psc}.
The PWC allows walks to skip lookups for some levels of page table by matching the index bits of each level of the page table node with those cached by previous page walks.
Intel's PWC is organized in three depths of translation caching: L4, L3 and L2.
An L4 PWC holds previous walk paths that share the top 9 bit virtual address,
allowing the walker to skip accessing the L4 page table entry, and go directly to the
L3 page table entry.
As each L4 entry covers 512\GB of virtual address space, this means that accesses that stay within a 512\GB virtual address range will hit in the PWC and be able to skip the L4 lookup.
With an L2 PWC, a walk that matches all upper 27 bits of the virtual address
will be able to skip the first three levels of the page table, and directly
access the level 1 page table node.
Such L2 PWC hits enable single-access translations (only a L1 entry access is required) for TLB misses within 2\MB regions of virtual address space.

Thus, the PWC enables page walks to skip one, two or
three levels of the page walk depending on the locality of the virtual address.
The impact of the PWC is shown in \autoref{fig:native_avg_mem_access}, baseline. The PWC enable skipping an average of 2.2 to 2.9 of the naive 4-access page walker memory accesses for all benchmarks except
gups and random access. Those two benchmarks are an exception, as they exhibit a highly random
access pattern across a large enough virtual address range that the PWC is only able to skip one level (one memory access) per page walk on average.

As a two-level flattened page table consists of a root level with up to $2^{18}$ page table entries, the L3 PWC can match the top 18 bits of the virtual address.
A PWC hit will then skip the first level of the flattened page
table (the L4+L3 entry) and directly access the flattened L2+L1 page table entry,
resulting in a single memory access for each TLB miss.
Our evaluations (\autoref{sec:eval}) confirm that the flattened page table augmented with a PWC sees an average of one memory access per page walk (\autoref{fig:native_avg_mem_access}).
It is worth noting that merging the first two levels of the page table (L4+L3) may make the PWC more effective as each PWC hit translates twice as many index bits and fewer PWCs are required since there are fewer levels, enabling each one to cache more entries.

\subsection{Supporting large data pages}
\label{sec:flat:lp}
If the lower two levels (L2+L1) are flattened, then there is no way to use an L2 entry to directly provide a 2\MB translation.
This requires 512 replicated translations in the L2+L1 node all pointing to the same 2\MB page.
While this may be helpful for sparse accesses (since it enables only two accesses for the page walk) it puts pressure on the cache if many of those replicated entries are accessed.

\begin{figure}[tb]
\includegraphics[width=\columnwidth]{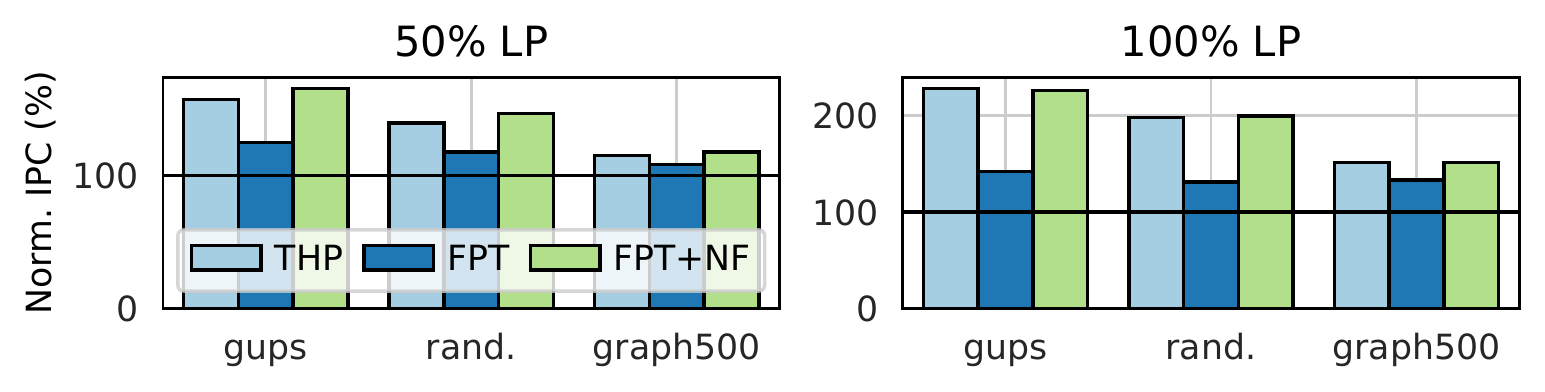}
	\caption{Impact of avoiding redundant page table entries for large pages (FPT+NF vs. FPT) for 50\% and 100\% large pages. Normalized to 0\% large pages.
	}
\label{fig:fpt_nf_graph}
\end{figure}

\autoref{fig:fpt_nf_graph} shows the resulting loss in performance with flattening (FPT, middle dark blue) compared to a traditional page table (THP, left light blue) for two fragmentation scenarios: 50\% large pages (realistic~\cite{perforated_pages}) and 100\% (performance limit, but unrealistic).
To mitigate this, we take advantage of the flexibility to flatten different parts of the page table differently.
Specifically, we allocate/promote 2\MB data pages in 1\GB virtual address regions and mark those regions to not have their L2 and L1 tables flattened (\autoref{fig:large_page_flattened} left, bottom mapping).
This allows allocations in the 1GB regions to behave as L4+L3 flattened tables (\autoref{fig:large_page_flattened} center) while other regions also have their L2+L1 levels flattened.
Page walks to 4\kB mappings in this region will require up to 3 memory accesses, but 2\MB mappings will only require up to 2 accesses and no replicated entries.
With this optimization (FPT+NF in \autoref{fig:fpt_nf_graph}), flattening surpasses the baseline by providing the benefits of L4+L3 flattening with efficient large page access.

The OS can decide how to flatten the page table, for example, based on mapping statistics gathered for page promotion or hints from the application.
In our L4+L3 and L2+L1 flattened simulations, we heuristically mark a 1\GB region to not have L2+L1 flattened if there are 32 or more 2\MB pages in it, but this threshold can be dynamically adjusted by the kernel.
Applications with many 2\MB pages may benefit instead from flattening the L3+L2 levels (\autoref{fig:large_page_flattened}, right), enabling the PWC to skip most L4 accesses, providing single-access  page walks to 2\MB pages and two accesses for 4\kB pages.

In order to use the merged page tables, the encoding of the VA bits that are used for the indexing of the PTs needs to be adjusted. Traditionally for a 4-level tree structure every page table has $2^9 = 512$ entries. This means that the VA is decoded using 9 bits to index into the page tables and 12 bits which are used for the offset once the translation is complete, and the page has been retrieved.

When merging tables, each combined table structure consists of $2^9\times2^9 = 262144$ entries fitting into a 2MiB page. In this case, each merged table requires 18 bits for indexing. As shown in \autoref{fig:largetablewalk} (left) this still requires the same total number of VA bits in order to traverse the page table structure and recover the physical address.

\subsection{Accessing recursively mapped page tables}
\begin{figure*}[tb]
\centering
\includegraphics[width=1\linewidth]{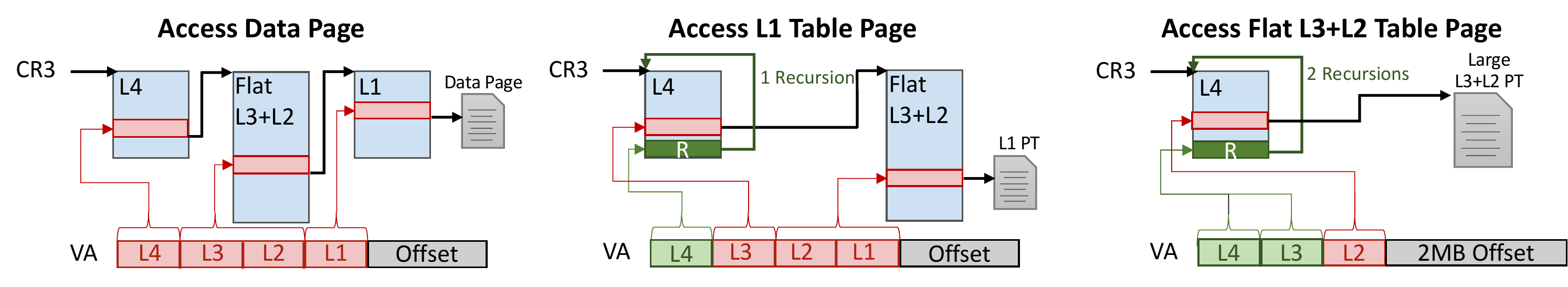}
	\caption{Recursive mappings used to access page tables in a L4, L3+L2, L1 page table
	organization.}
\label{fig:largetablewalk}
\end{figure*}
To manipulate page tables Linux uses a mapping of all physical memory to a contiguous virtual address range in kernel space and a software page table walk.
However, Windows uses recursive page tables where the page tables map themselves into their own address space.
This is done by having a special recursion entry in the top-level (L4) node that points back to the node itself, and using recursions through this node to prevent the page walker from reaching the ordinary leaf node.
As a result, the walk returns the address of a the page table node where it stops, and not the normal data translation/frame.

The degree of recursion can be controlled by the number of index fields in the VA that are filled with the recursion entry index.
Recursive access to the page table for a 4-level table with the middle two levels flattened (L4, L3+L2, L1), is shown in \autoref{fig:largetablewalk} with a normal translation of a data page to the left.

If the top 9 bits of the VA are the recursion index (\autoref{fig:largetablewalk}, middle), the 3-step page walk will recurse once and step through the top (L4) node twice: L4 $\rightarrow$ L4 $\rightarrow$ L3+L2. 
The final VA bits will then return the address of the L1 page table node indexed from the entry in the L3+L2 node. 
In a similar manner, if the top 18 bits of the VA have the recursion index concatenated twice, then the page walker will recurse twice (\autoref{fig:largetablewalk}, right), and step through the top (L4) node three times: L4 $\rightarrow$ L4 $\rightarrow$ L4, returning the address of the L3+L2 page table node indexed by the L4 node.

Since the address encoding is the same for data translations and pointers to page table nodes, the same page walk can return either data translations (with no recursions) or the addresses of the page table nodes themselves (by adding recursions, which causes the page walk to end earlier on a page table node). 
The more recursions, the higher the node level returned, and the specific node is selected by the remaining VA bits.

\begin{figure}[t]
\includegraphics[width=1.0\columnwidth]{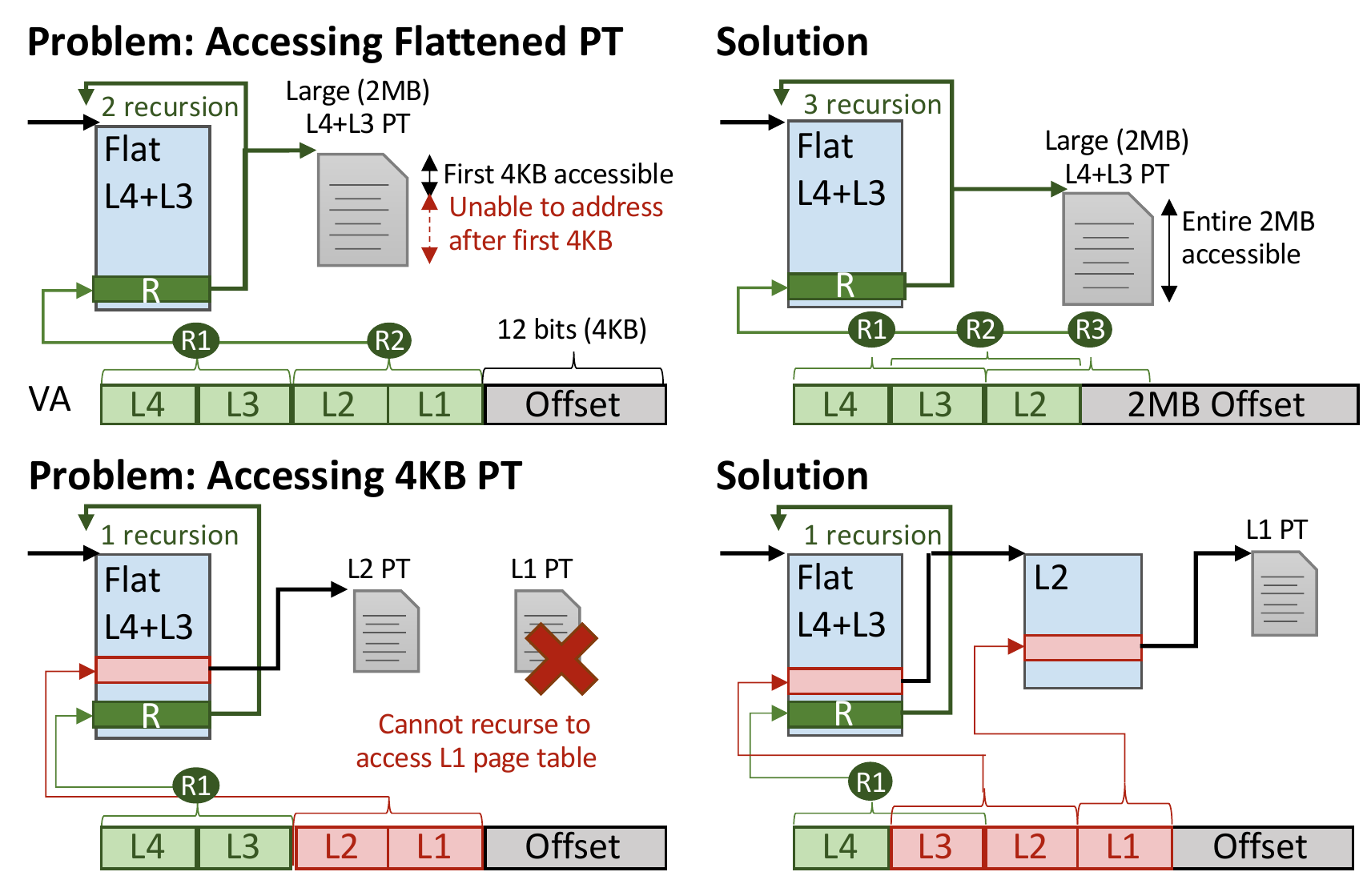}
\caption{Flattened L4+L3 causes two problems. Our solution is overlapping index bits.}
\label{fig:half-step}
\end{figure}

Traditional page walks terminate and return a large page translation when they encounter an entry marked as a large page (e.g., a 1\GB page is returned if marked at L3 or a 2\MB if at L2).
To make recursive page table walks work with flattened page table nodes, the walker is modified to recognize pointers to flattened page tables as large page mappings (in addition to normal 2\MB mappings) while looking up L2 entries (\autoref{fig:largetablewalk}, right).
In addition, the page walker cannot simply use 18 bits of the VA to index recursively into flattened nodes.
For example, accessing the L4+L3 node in a flattened L4+L3, L2, L1 page table (\autoref{fig:half-step})  requires two recursions of 18 bits each for the L4+L3 2\MB node (top left), leaving insufficient bits for the final indexing of the node.
Similarly, it is not possible to reach the L1 node in this situation (bottom left), as even a single recursion will overshoot in address bits.

To address this, we overlap index bits when following a self-referencing node pointer in a large node. 
As \autoref{fig:half-step} (right) shows, we use 18 bits of the VA for indexing into the large self-referencing node, but then we only advance through the VA by 9 bits instead of 18. 
The next step in the walk then reuses the lower 9 bits of the index from the previous step as its higher 9 bits (\autoref{fig:half-step}, top right L3 and L2 and bottom right L3).
This implies that we replicate the self-referencing entry in the large node 512 times, since we need to ignore the lower 9 bits when recursing as they form the upper 9 bits of the next step.
While this requires 512 entries in the flattened 2\MB node, it corresponds to the same amount of address space as a single recursion entry in a non-flattened L4 node. With this modification we can enable flattening in self-referencing recursive page tables.

\subsection{Implications for five-level page tables}

Unless programs spread their data across the full address space, 5-level page tables~\cite{intel:5_level} will behave similarly to 4-level page tables with the PWC caching translations for essentially all 5th-level translations.
However, their larger address space opens up more possibilities for flattening.
In particular, merging the L5+L4 levels and the L3+L2 levels and directly translating 2\MB pages or using an L1 level when 4\kB pages are needed would be quite attractive.
It might even be desirable to use a 1\GB page by merging three levels, if the kernel can reliably allocate such regions.


\section{Flattening and Virtualization}
\label{sec:virt}

\begin{figure}[t]
\centering
\subfloat[2-D page table\label{fig:virt:g4h4}]{\includegraphics[width=0.44\columnwidth,valign=b]{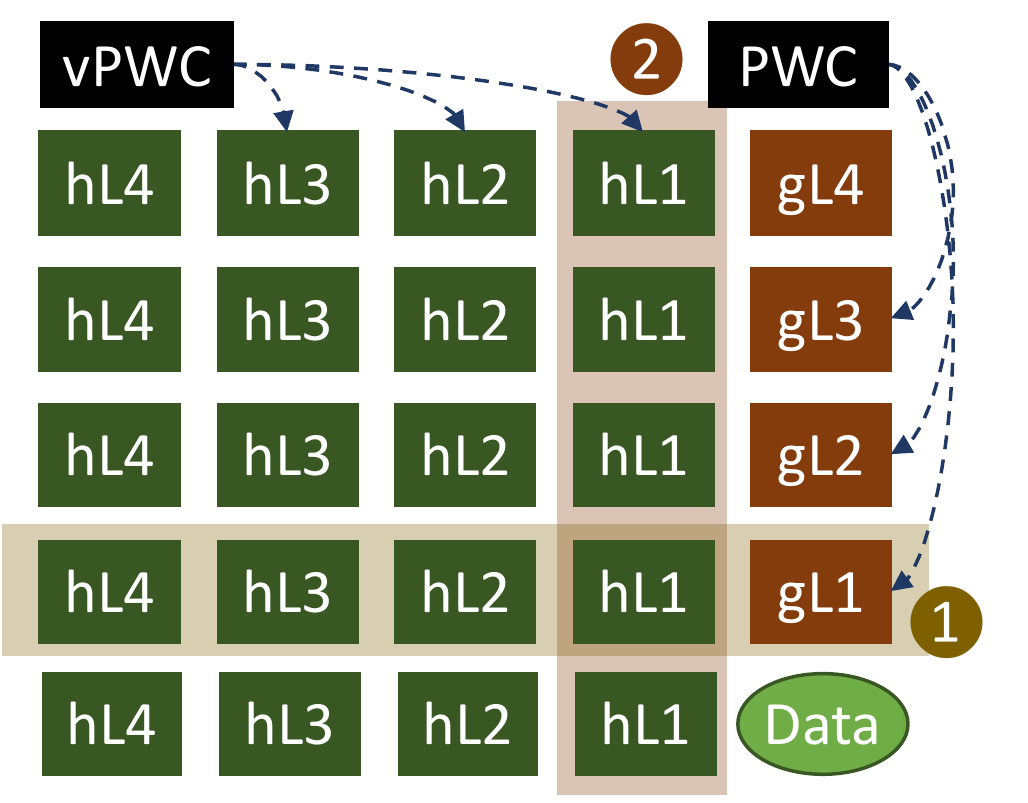}}
\subfloat[Flattened host and guest\label{fig:virt:gFhF}]{\includegraphics[width=0.505\columnwidth,valign=b]{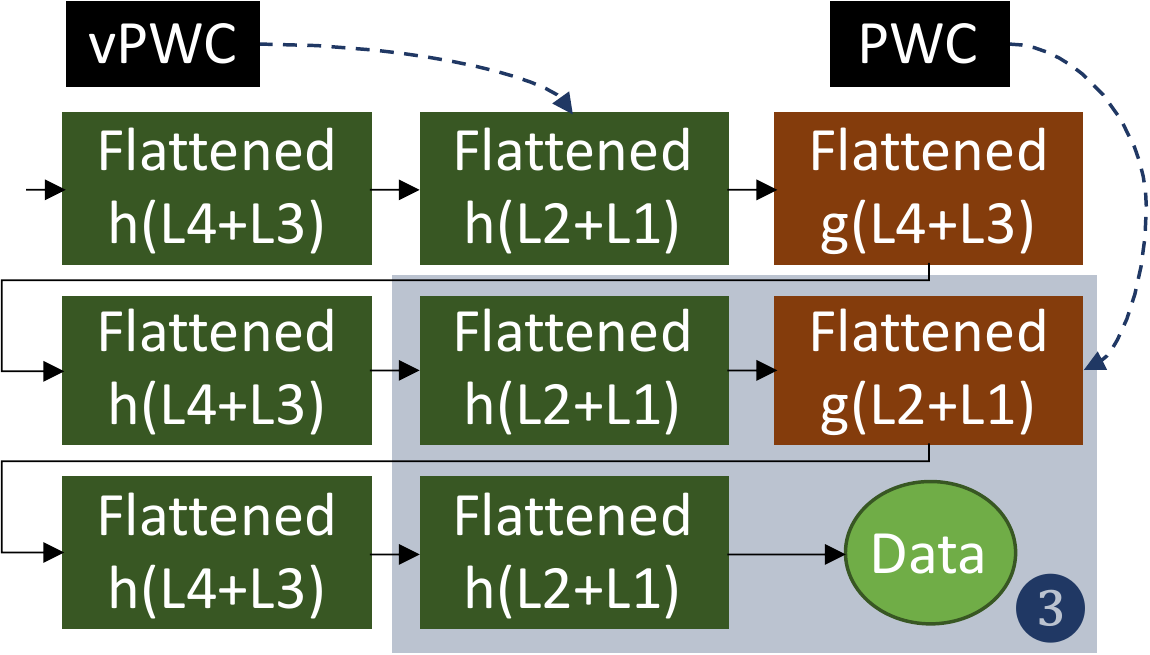}}
\caption{Flattening in virtualized 2-D page table walks. The PWC and vPWC skip stages
of the virtualized walks.}
\label{fig:virt}
\end{figure}

With virtualization, guests use a set of page tables to map from the guest virtual address (gVA) to the
guest physical address (gPA), and the hypervisor uses a second set to map from the guest physical address to the host physical address (hPA).
A guest translation therefore needs a two-dimensional page walk (\autoref{fig:virt:g4h4}) in which
each of the four guest page table level access first requires its own four-access host page table
walk, and a final four-access host page table walk is required to translate the final guest physical
address, incurring a total of 24 memory accesses.
As there are two page tables involved, there are three possible combinations of page table flattening.
Either the host or guest page tables can be flattened (14 accesses), or both (8 accesses), can be flattened. (See \autoref{fig:virt:gFhF}.) 

\subsection{Effective memory accesses per page walk}
\label{sec:virt:cache}
Although virtualization naively incurs 24 memory accesses per walk, in practice
many of these accesses are eliminated by a combination of three techniques (\autoref{fig:virt:g4h4}): PWCs, large pages, and host translations in the TLB. 
\autoref{fig:virt:g4h4} shows how the PWC skips levels of the guest (5th column, skip downwards) walk and a vPWC skips them for the host walk (applicable on all five rows).
Our evaluations show that this reduces the number of accesses to as low as 3 to 4.8.
Even the two most random access applications, GUPS and random access, require only 9.6 and 9.4 memory accesses.

Hypervisors also try to map the guest physical pages to host physical pages in large 
pages to make host translations more efficient~\cite{kvm_thp, xen_thp}.
Using large pages comes with the benefit of removing the last level of the
conventional four-level page table. In a 2-D page table, a 2\MB guest mapping can
remove a row of memory accesses (\ding{202} in
\autoref{fig:virt:g4h4}). Large page mappings for the gPA to hPA mapping
remove columns (\ding{203}).

With host and guest page tables flattened, the number of accesses
per page walk is reduced to 8, and, with the help of
the PWC, this comes down to 4.4 in practice, as shown by the box (\ding{204} in
\autoref{fig:virt:gFhF}). 
The guest PWC allows the page walk to skip the flattened g(L4+L3),
and directly access the flattened g(L2+L1). To access the g(L2+L1), the host
page table is traversed, and here the vPWC skips the flattened h(L4+L3),
and directly accesses the flattened h(L2+L1). After the gPA of the data is
resolved, the host page table is traversed once more for the actual data gPA to
hPA translation, resulting in another walk, where the vPWC skips the
h(L4+L3), enabling a single access final translation. Finally, after 2.8 
accesses, the actual data can be accessed.


\section{Cache Prioritization}
\label{sec:prio}
The other half of our approach is to reduce the latency (and energy) of each page table access by increasing the chances that it will be a cache hit.
We do so by biasing the cache replacement policy to keep page table entries when an application is experiencing high TLB miss rates.
Unlike CSALT~\cite{csalt}, which biases the cache to store TLB entries to support their DRAM-TLB-cache design, we bias the cache to store page table entries for the existing page table walker.

Biasing the replacement policy to favor page table entries means evicting more data, but we find that applications with high TLB miss rates also exhibit high data miss rates (L2 and L3 data miss ratios of 95\% and 80\%). 
This, combined with the page table access being on the critical path to the data access, suggests that allocating more cache space to the (much smaller) page table over the data itself is likely to be more beneficial than caching the data\footnote{MASK~\cite{mask} identified that the \textit{opposite} approach of prioritizing data over page table entries is better for latency-insensitive GPUs.}.

As data sets grow in the future, the likelihood of hitting in the cache will always remain  higher for the page table than for the data for applications with limited locality simply due to the disparity in size between the two.
Concretely, an application with 8\GB of densely allocated memory requires $2^{21}$ 4\kB pages represented by 8\,B each, for a total of 16\MB of space.
In the conventional four-level page table, each intermediate level is $1/512$ the size of the previous one, resulting in only 32\kB for the L2 level, which is likely to be skipped by the PWC.
From this analysis we see that an 8\GB memory workload could fit all leaf page table entries into a 16\MB cache, and is thus practical to cache in today's systems.
Even with a random access pattern, caching even a portion of the page table entries would deliver a significant benefit.
Future, larger workloads might even benefit from prioritizing among different levels of the page table itself, although we have not explored this.

In addition to our simulations, we explored the potential of preferentially caching page tables on current system by created a thread that runs on another core and periodically touches the page table of a target application to keep it in the shared LLC.
We ran this thread together with graph500 (scale 24) on an Intel i7-9700 with a
shared 12\MB LLC, and found a 5\% performance increase from keeping the entire page table in the LLC (miss ratio of 0\% for the page table thread).
While this experiment both uses extra LLC bandwidth and does not bring the page table into the target's private L2, unlike our proposed and simulated design, it does demonstrate that preferentially caching page tables is a promising approach.
This leads to our conclusion that if the program accesses the memory in a way that does not make good use of the TLB and the caches for data, then we would be better off prioritizing page table entries in the caches.


\section{Implementation}
\subsection{Hardware changes}
\label{sec:impl:hw}
Flattened page tables require augmenting the hardware page table walkers to be aware of the size of the page used at each level of the page table.
This requires two additional bits (for 4\kB, 2\MB, and 1\GB pages, or one for just 4\kB and 2\MB) in
the CR3 register (for the root node) and at each entry in the page table, possibly in the currently unused bits. 
These bits indicate the size of the page at the next level of the page table, and are needed to determine how many bits of the virtual address to use as index bits in the walk.
As page table nodes are aligned, using flattened page tables frees up 9 or 18 bits  (2\MB~or 1\GB~nodes) in the page table entries that point to flattened page table nodes, leading to more available bits.
Similar small changes are required to the PWC, but there is the potential to use storage more effectively if there are fewer levels in the page table.
Overall these changes are minor and incur essentially no hardware overhead.

Cache prioritization requires detecting phases of high cache
and TLB miss and enforcing the prioritization. 
Detection is easily accomplished using existing hardware counters.
Prioritization can be accomplished with a range of techniques.
For example, existing cache partitioning technique (such as way-partitioning) can allow the OS to pre-load page table entries into part of the cache that will not contend with the application's own data.
Alternatively, a per-cacheline tag bit can indicate if the entry is a page table and then bias the replacement policy away from such lines.
The cost of such hardware is less than 0.2\% of the cache size, and is already present in server-class processors that support per-context cache partitioning~\cite{arm_mpam, intel:sdm_vol3}.
Indeed, Arm's MPAM already stores a partition ID that can be used to differentiate processes or even I/D cache lines~\cite{arm_mpam}.
We use this approach for prioritization in the L2 and LLC during phases of high TLB miss rates: when choosing
a victim for replacement, 99\% of the time we choose to evict data over page table entries.
If there are no data entries in the set, or in the other 1\% of the evictions, we evict the LRU entry.
We empirically found that this ratio works well.

To limit impact on co-runners in shared caches, prioritization can occur within a context's (core/process) allocation by using identifiers in the tags~\cite{intel:sdm_vol3,arm_mpam}.
For our multicore simulations we used this approach to prevent one process' data from evicting
another's page table.

As flattened page tables require only trivial hardware changes, the energy benefits will be proportional to the reduction in memory system accesses and execution times.

\subsection{Software changes}
\label{sec:impl:sw}
We have a working operating system prototype based on Linux 5.8.13 running on an industrial Armv8 functional simulator
and on existing HW by adding an additional shim level before the large page tables. The change to the kernel is small: +614/-109 lines.
Page tables are automatically flattened if a sufficiently large allocation can be provided by the kernel.
Our implementation flattens L3+L2 (\autoref{fig:large_page_flattened}, right). L5+L4 flattening could be added with only minimal changes.
We have not implemented dynamically flattening page table levels after allocation.
However, this is straight-forward by allocating a large page and copying the page table entries of the lower node (L2 when flattening L3+L2)
into the new flattened node. The upper node (L4) entry can then be updated to point to the flattened node.

Most Linux's page table management is shared between architectures.
The kernel internally assumes that all architectures implement a 5-level radix tree where the highest level, L5 (PGD using Linux's terminology), is always implemented.
This avoids special cases since unimplemented levels can be \textit{folded} into the parent level using a virtual entry that points back to an entry in the parent table.
For example, in a system with a three level page table,
the kernel would implement L5 (PGD), L2 (PMD), and L1 (PTE), with L3 (PUD) and L4 (P4D) folded into the L5 table.
Internally, the kernel treats L3 and L4 as having a single virtual entry (effectively the corresponding L5 entry) and no storage.

Using this existing support, we can readily implement support for flattened L2+L3 tables.
We simply fold the L3 table, request 2\MB instead of 4\kB when allocating an L2 table, and change the macros that define the bit ranges used to index into the tables.
However, this approach does not work in practice since it makes 2\MB pages a hard requirement, and
fragmented systems could fail page table allocations if they are unable to find a free 2\MB block.

To support a graceful fallback to 4\kB table nodes, we needed to be able to selectively fold the L2 and L3 tables per sub-tree of the page table.
The kernel currently only supports folding an entire level across a process' entire address space.
For our our prototype, we changed the signature of a handful of functions to add a mechanism to determine if a level needs to be folded based on the state of the parent entry.
Overall, this change is small (roughly 100 lines of shared code a few tens of lines per architecture) and mostly mechanical.

With selective sub-tree folding in place, we can allocate large L3 tables.
If we succeed in allocating a 2\MB page for a flattened L3+L2 table, we treat the table as an L2 table and fold the L3 table.
The L4 entry is configured to point to the new L2 node and a bit in the entry is set to indicate that the next node has been flattened.
If the flattened L3+L2 table allocation fails, we allocate normal 4\kB L3 and L2 tables.

We stress-tested our prototype kernel on a server system with 128 HW threads by building a Linux kernel with 100 concurrent processes.
Since the hardware does not support flattened tables, we allocate and manage the flattened table in the kernel, but inject a shim table between the L4 table and the flattened L3+L2 table to be compatible with the existing architecture.
We found that 0.5\% (20 out of 3464 compiler invocations)
failed at least one of the two 2\MB allocations needed for the flattened page table on a system with
6\% memory oversubscription (500\MB swap with 8\GB RAM). This increased to 12\% failures on a system
with 50\% oversubscription. When a 2\MB allocation failed, the prototype used our fall-back path
with traditional 4\kB tables.


\section{Evaluation}
\label{sec:eval}

\begin{table}[t]
\footnotesize
\centering
\caption{Server simulation configurations}
\label{tab:methodology}
\begin{tabular}{@{}ll@{}}
\toprule
Processor    & 2\,GHz Out-of-order x86 Processor                                                                                                           \\
L1 I/D Cache & 4-cycle, 32\kB, 8-way, 64\,B block                                                                                                           \\
L2 Cache     & 12-cycle, 256\kB, 8-way, 64\,B block                                                                                                           \\
L3 Cache     & 42-cycle, 16\MB, 8-way, 64\,B block                                                                                                           \\
Memory       & DDR4-2400, 4 channels                                                                                                                       \\ \midrule
L1 TLB (Parallel lookup)       & 
  \begin{tabular}[c]{@{}l@{}}
    4\kB: 1-cycle, 64-entry , 4-way \\
    2\MB: 1-cycle, 32-entry, 4-way
  \end{tabular}\\
L2 TLB       & 4KB/2MB: 9-cycle, 1\,536-entry, 12-way 					 \\ \midrule
\begin{tabular}[c]{@{}l@{}}PWC (Parallel lookup) \end{tabular}	&
\begin{tabular}[c]{@{}l@{}} L4:1-cycle, 4-entry FA \\ L3:1-cycle, 4-entry FA \\ L2:1-cycle, 24-entry FA
\end{tabular}					 \\
Nested TLB~\cite{2d_translation}       & 1-cycle 16-entry FA \\ \bottomrule
\end{tabular}
\end{table}

\begin{figure*}[t]
\includegraphics[width=\textwidth]{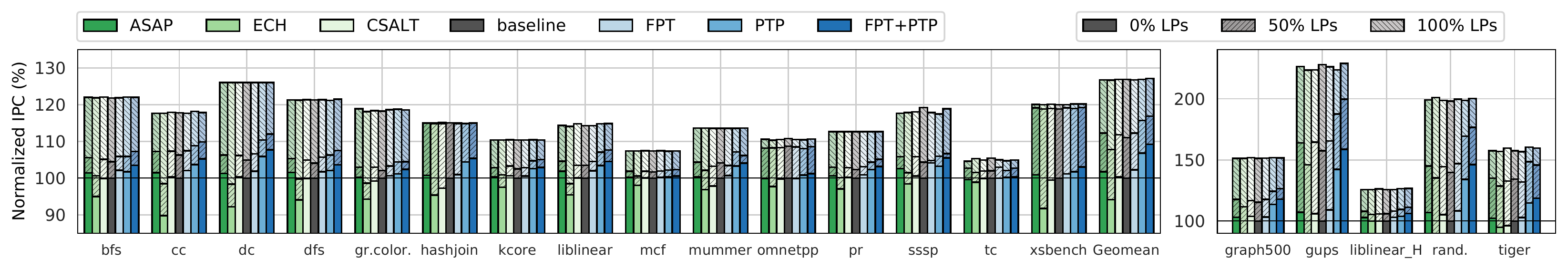}
	\caption{Performance of flattening (FPT), cache prioritization (PTP), both combined (FTP+PTP), and the related work (ASAP, ECH, CSALT). Three fragmentation scenarios are shown stacked: 0\% large pages (bottom, all 4\kB pages), 50\% large pages (middle, realistic), and 100\% large pages (top, unrealistic for actual systems). Normalized to the PWC-equipped baseline (solid line and dark middle bar) with 0\% large pages.}
\label{fig:ipc_native}
\end{figure*}

We implemented our proposal in the gem5~\cite{gem5} simulator (\autoref{tab:methodology}).
We model the TLB~\cite{intel:opt} and the Paging Structure Caches (our PWC)~\cite{psc, reverse_psc} of the Intel Skylake microarchitecture. 
Unfortunately, the page walker cache provided in gem5 is modeled as a cache with 64\,B-lines, which does not accurately represent either page table caches nor translation
caches~\cite{skipdontwalk}.
We implemented our own PWC, which sends misses to the L1 data cache~\cite{intel:sdm_vol3}, as in Intel processors.
Our flattened page tables use the L2 PSC and we model a vPSC for the host page
table under virtualization.

This work requires comparing two different page table organizations.
To keep all other aspects of the simulation state the same, such as data layout in memory and kernel interrupts, we use system-call emulation (SE)
mode to vary the page table organization, while keeping all other states identical. 
SE is sufficient as we do not study the effect of changes in memory mapping or the
page table during the runtime of the program,
and focus our evaluation on the program execution itself.
The gem5 SE mode normally does not model a page table walker nor a page table in the form of a radix
tree.
We included page table walks by
constructing the page table based on the simulator VA to PA mappings, and having the page table walker appropriately access each level through the caches.

We evaluate three large page fragmentation scenarios~\cite{illuminator, ingens}: 0\% (all 4\kB pages; worst case for page walks), 50\% (typical real-world~\cite{quicksilver,perforated_pages}), and 100\% (all large pages; best performance but unrealistic).
For the 50\% scenario, we allocate large pages for the first half of the address space to simulate an OS that runs out of free large pages.

We evaluate our proposals on benchmarks that stress the TLB:
GraphBIG~\cite{GraphBIG} (LDBC-1000k dataset, 6.6\GB of memory usage) and graph500 (scale 24, 5.4\GB);
benchmarks with significant TLB misses from biobench~\cite{biobench} and
SPECCPU 2006~\cite{SPECCPU2006};
GUPS ($N=30$, 8\GB);
large linear classification (liblinear)~\cite{liblinear} (inputs: url\_combined and HIGGS);
a hashjoin microbenchmark~\cite{mitosis},
and the XSBench~\cite{xsbench}.
As large in-memory databases, such as memcached, and very large graphs exhibit random access patterns, we also include a microbenchmark that represents such random access behavior.

\subsection{Non-virtualized execution}
\begin{figure}[t]
\includegraphics[width=\columnwidth]{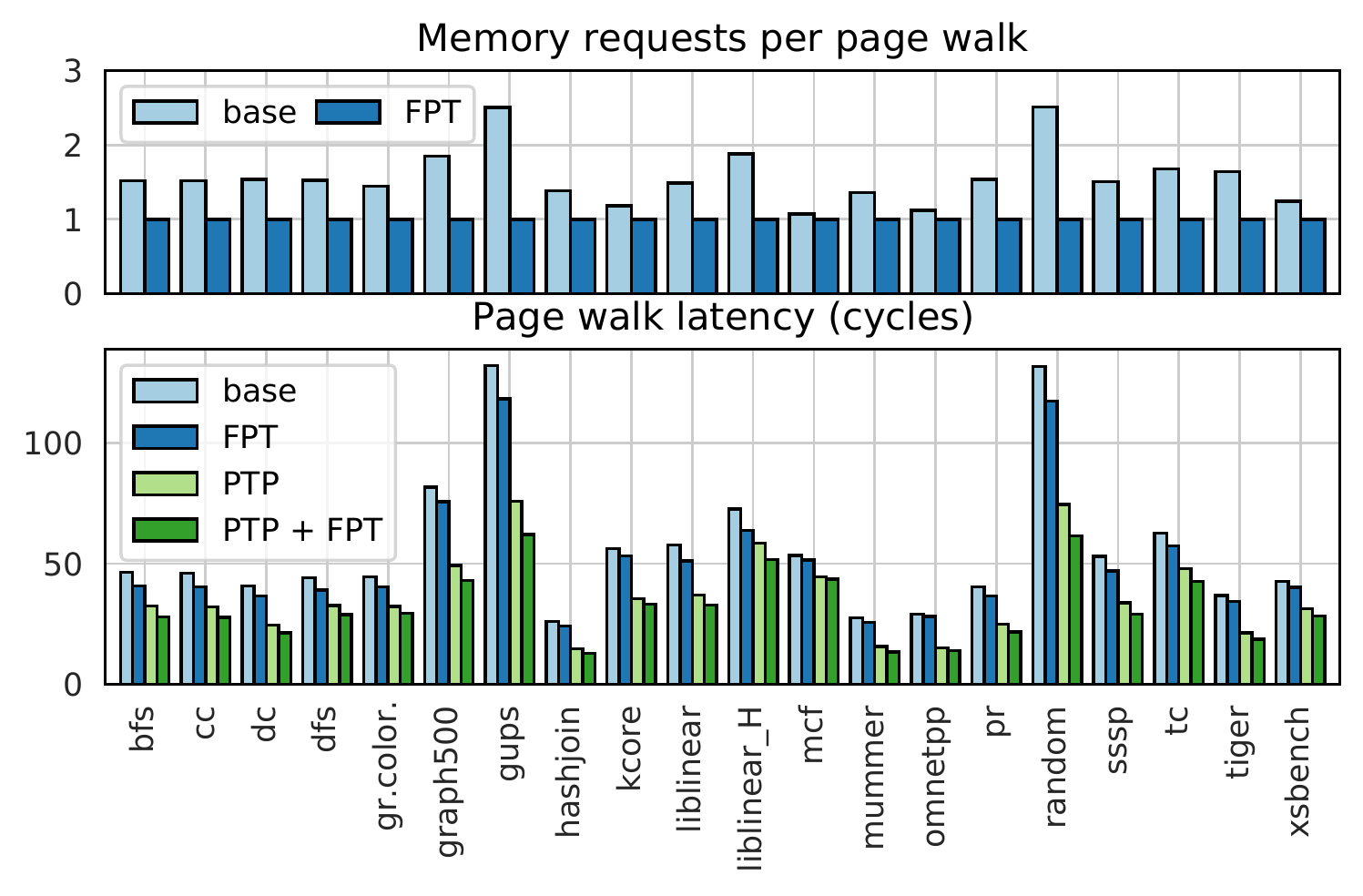}
	\caption{Memory accesses and latency per page walk across a traditional 4-level page table (baseline), flattened (FPT), and cache prioritization (PTP), all with PWCs.}
\label{fig:native_avg_mem_access}
\end{figure}

\autoref{fig:ipc_native} shows the performance of our approach and those of ASAP~\cite{asap}, Elastic Cuckoo Hashing~\cite{ECH}, and CSALT~\cite{csalt}.
The performance numbers are presented relative to a baseline system with a 4-level page table and Intel-style PWCs. 
We plot results for three fragmentation scenarios stacked to show the change as the percent of large pages increases: bottom (0\% large pages), middle (50\% large pages, realistic~\cite{perforated_pages}), and top (100\% large pages, best performance, but unrealistic). 
Performance is normalized to the baseline configuration with 0\% large pages (horizontal black line/dark gray bar), and the effect of 50\% and 100\% large pages can be seen on the baseline system in the gray bar in the middle.

We see two trends in the results:
First, for the 0\% (bottom) and 50\% (middle) large page cases, flattening the page table (FPT), prioritizing page table entries in the cache (PTP), and the combination (FPT+PTP) contribute to increasing performance improvements beyond the state-of-the-art (blue bars increasing to the right).
Second, as the fragmentation decreases (stacked from bottom to top), the impact of flattening and prioritization also decreases. 
This is because the TLB misses are much less frequent (less opportunity to reduce latency) and the page table size itself is drastically smaller (less need to preferentially cache it).
Interestingly, the combination (FTP+PTP) is almost as effective as moving from 0\% to 50\% large pages (9.2\% vs. 11.0\%).
For the detailed analysis below we look at the 0\% large page scenario as flattening and prioritization have the largest potential\footnote{We expect similar effects for 2\MB pages in the future if TLB reach does not increase as fast as data sizes.}.
\autoref{fig:native_avg_mem_access} shows that the flattened page table together with the PWC results in single-access page walks for all workloads.
Overall, flattening the page table for 4\kB pages shows a geometric mean performance improvement of 2.3\% (solid light blue bar) vs. 1.7\% for ASAP (solid dark green bar) and a net performance loss for ECH (solid medium green bar).

Prioritizing page table entries in the L2 and L3 caches on the baseline 
improves performance by 6.8\%.
This does increase data misses slightly (L2: +4.7 percentage points) in exchange for far fewer page walk misses (L2:
-36.2 percentage points).
\autoref{fig:native_avg_mem_access} (bottom) shows that this prioritization significantly reduces page walk
latency from an average 50.9 cycles per walk (baseline) down to 33.0 (prioritizing). Flattening and
prioritizing together results in 29.1 cycles per walk.

ECH shows lower performance than the baseline for all 0\%, 50\%, and 100\% large page scenarios
(-5.9, -3.2, -0.2 percentage points) as it is requires three (4\kB pages) or four (mixed 4\kB/2\MB
pages) concurrent memory accesses vs. a single memory access with flattening.
CSALT provides little benefit for the 0\% large page scenario.
We believe this is due to CSALT having been designed and evaluated only with large pages,
which makes it poorly optimized for the much larger page tables from our 0\% and 50\% scenarios,
and their assumption of very frequent (every 10ms) context switches, which would make a PWC less effective.

Changing PWC size resulted in a performance impact of -1.5\% to +2.4\%, when sweeping the L3 PWC entries from 1 to 16 (baseline 4) for the most sensitive benchmark, GUPS.
In comparison, flattening gave a benefit of 8.9\% as it benefits from a single memory access, instead of the 2 memory access from a L3 PWC hit.
Achieving a similar benefit to flattening (single memory access) would require increasing the L2 PWC entries to approximately 4096.
The larger 16-entry L3 PWC provided +2.9\% on top of our cache prioritization.

To summarize, for 0\% large pages, flattening the page table improved performance by 2.3\% over the baseline. 
Prioritizing caching of the page table resulted in a 6.8\% improvement, while the combination delivered 9.2\%, which is significantly greater than the state of the art (bottom bars in \autoref{fig:ipc_native}: ASAP 1.7\%, ECH -5.9\%, CSALT 0.3\%).
However, as the proportion of large pages increases, the relative improvement decreases.
For the 50\% large page scenario, the combination of flattening and prioritization delivered a 5.8 percentage point improvement vs. ASAP 1.2, ECH -3.2, CSALT 0.7 (middle bars in \autoref{fig:ipc_native}).

\begin{table}[]
\centering
\footnotesize
\caption{Benchmark mixes for the multicore evaluation.}
\label{tab:multicore_mixes}
\begin{tabular}{@{}rl|rl@{}}
\toprule
Mix & Benchmarks                         & Mix & Benchmarks                           \\ \midrule
1   & dc$\times$4                        & 2   & liblinear\_H$\times$4                \\
3   & rand$\times$2, dc$\times$2         & 4   & rand$\times$2, hashjoin$\times$2     \\
5   & hashjoin$\times$2, mummer$\times$2 & 6   & liblinear$\times$2, xsbench$\times$2 \\
7   & tiger$\times$2, dfs, bfs           & 8   & rand, liblinear, dc, cc              \\ \bottomrule
\end{tabular}
\end{table}

\begin{figure}[t]
\includegraphics[width=\columnwidth]{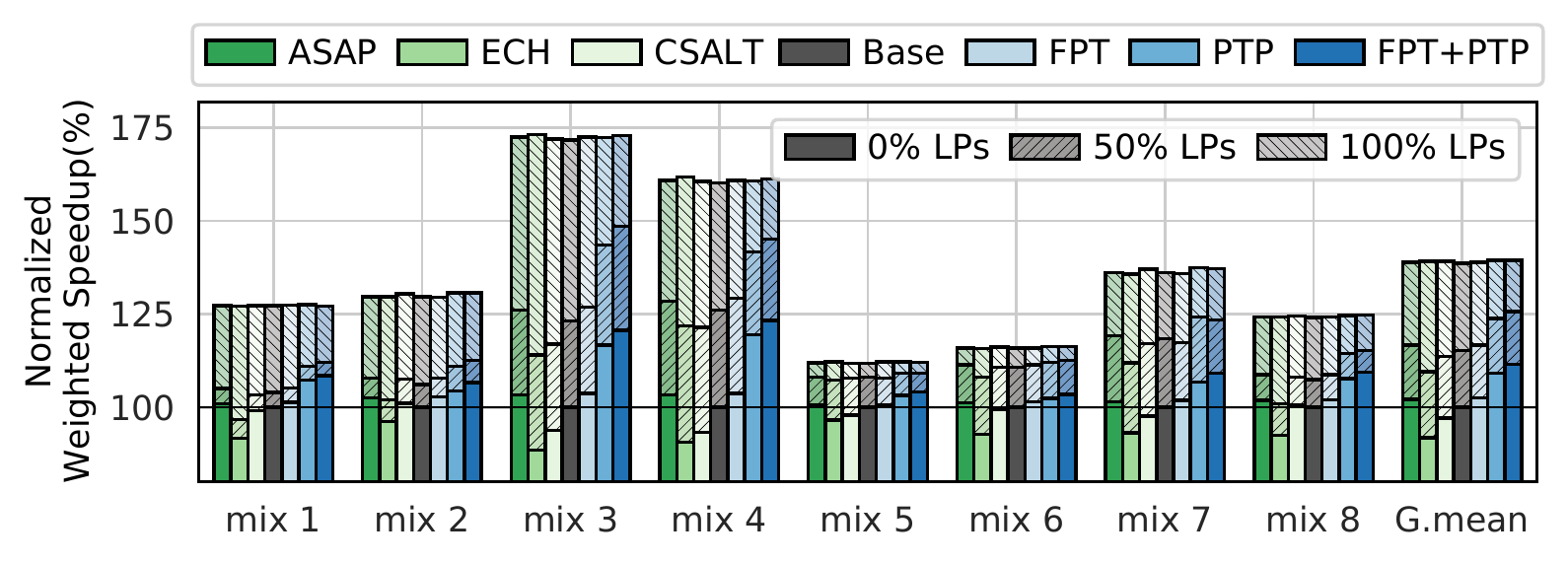}
\caption{Multicore performance. Mean for all 20 mixes.}
\label{fig:speedup_multicore}
\end{figure}

\begin{figure*}[t]
\includegraphics[width=\textwidth]{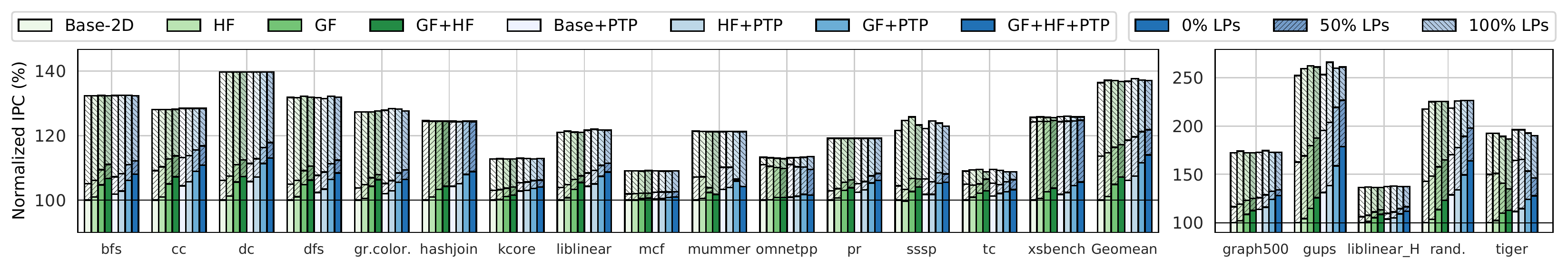}
	\caption{IPC comparison in virtualized environments of various combinations of flattening the page
table for the host (HF) and guest (GF) and both (HF+GF), with (blue) and without (green) cache prioritization  (PTP).}
\label{fig:ipc_virt}
\end{figure*}

\textbf{Multicore.}
To evaluate the effect on shared caches and the memory hierarchy, we evaluated flattening and prioritizing with multi-programmed workloads (32MB shared L3, 4 cores with private L1s and L2s).
We evaluated a total of 20 workloads: 11 homogeneous and 9 heterogeneous. 
\autoref{fig:speedup_multicore} shows the normalized weighted speedup for 8 mixes and the geometric mean of all 20.
The first two entries show that homogeneous mixes behave similarly to the individual benchmarks in~\autoref{fig:ipc_native}. This was consistent across all 11 homogeneous mixes.
The heterogeneous workloads show a similar performance improvement trend to the individual benchmarks: Flattening and prioritizing each introduce performance improvements,
and work together resulting in an average of 2.2\%, 9.2\% and 11.5\% improvement, respectively, for the 0\% LP scenario.
The improvements are 1.4, 8.6 and 10.3 percentage points, respectively, for the 50\%, and 0.2, 0.7 and
0.8 percentage points, respectively for the 100\% scenario.

\subsection{Virtualized executions}
\label{sec:eval:virt}

The performance benefits for virtualized systems are shown in \autoref{fig:ipc_virt}.
The green bars show the effects of flattening the host page tables (HF), the guest page tables (GF), or both (GF+HF). 
The blue bars include cache prioritization.
The baseline (Base-2D, leftmost) is a virtualized system with the 2D 4-level page table, which naively
incurs up to 24 accesses per page walk, but, because we include two sets of PWCs for the guest and
the host and a nested TLB~\cite{2d_translation} to hold host translations, the average number of
accesses is only 4.4. 

Flattening in virtualized environments delivered a
larger performance improvement than native environments.
Flattening the host page table alone resulted in a 1.1\% performance improvement, while
flattening the guest page table alone delivered 4.9\% improvement.

To understand the difference, consider Graph500, which which requires 11\MB~of guest and 22\kB~of corresponding host page tables for its 5.4\GB~gVA.
The Nested TLB~\cite{2d_translation} and the vPWC (host PWC) work together to efficiently cache the
small host translations (22\kB) for the guest page tables.
As a result, the host translations for the guest page table accesses are effectively single access, even for the baseline 2D walks.
This means there is less benefit for flattening the host page table for the guest page table accesses.
Indeed, most of the benefit for host flattening comes from the final host data address translation.
Guest flattening, however, allows flattening the 11\MB page table, resulting in fewer guest
accesses (which also leads to fewer host table walks).

Finally, flattening both page tables is the most effective. 
Flattening both host and guest page tables delivered the largest performance improvement of 7.1\%.
Adding page table prioritization in the cache, increased this significantly to 7.5\%, 11.6\%, and 14.0\%
performance improvements for flattening host, guest, and both, respectively. We found that page table prioritization in the
cache consistently provided a 6.1 to 6.9 percentage point improvement, which is similar to the native results.
The effects of large pages are similar to the non-virtualized executions.

\subsection{Dynamic Energy}
\begin{figure}[t]
\includegraphics[width=\columnwidth]{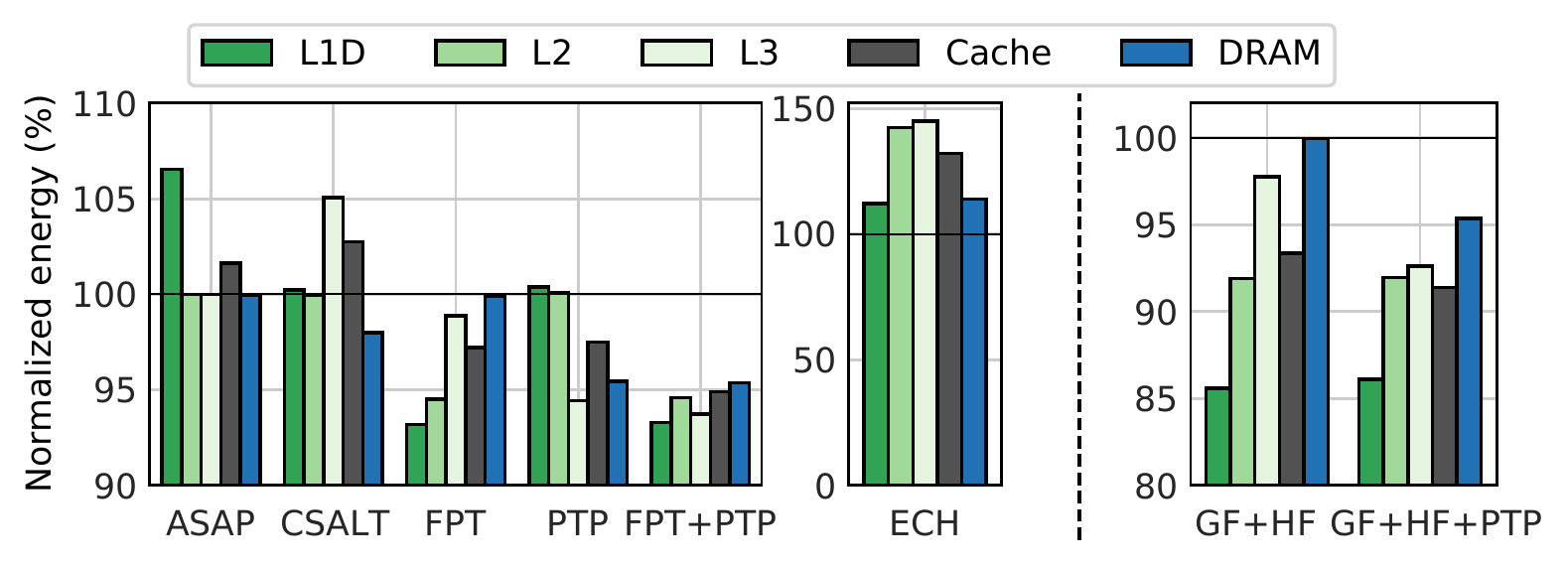}
	\caption{Dynamic energy consumption of the cache hierarchy and DRAM for data and page walks for native (left, center) and virtualized (right) executions.
	Normalized to respective baselines.}
\label{fig:power}
\end{figure}

We present dynamic energy normalized to the baseline in \autoref{fig:power}.
The cache energy is comprised of the L1D, L2 and L3 modeled with the CACTI~\cite{cacti} at 22nm.
We include both data and page table walks.
For DRAM, we report relative off-chip accesses.

ASAP which issues prefetches into the cache hierarchy for the lower two levels needs to re-access
the lower two levels resulting in higher L1D accesses.
CSALT does reduce off-chip DRAM accesses by 2.0\% but increases the L3 access 5\% resulting in 2.7\%
higher cache energy consumption.
ECH issues three accesses per walk for 4\kB pages resulting in higher cache (32\%) and memory (14\%) energy consumption.
This is a different behavior from the baseline, ASAP, CSALT and our work, all of which benefit from
fewer page walk accesses due to the PWC.

Flattening reduces the number of memory accesses to the
cache hierarchy (-2.8\%). Cache prioritization increases L2 hits and reduces
accesses to the L3 and the DRAM, reducing cache hierarchy energy (-2.5\%) and
DRAM accesses (-4.6\%). Finally the combination of both results in a dynamic energy reduction of 5.1\% and 4.7\%, for
cache and DRAM, respectively. We see a similar trend in virtualization with flattening both guest
and host table reducing cache energy by 6.7\% and adding prioritization resulting in 8.7\% cache and
4.7\% DRAM energy saving.

\subsection{Case Study: Flattening for mobile systems}
We evaluated flattening alone on a production industrial simulator used for
next-generation mobile core exploration, configured as a high-end mobile device (\autoref{tab:mobconf}).
We used the Speedometer 2.0~\cite{speedometer2} benchmark, which tests common browser operations such as DOM APIs, JavaScript, CSS resolution, and layout. It is a good representation of real-world mobile system performance, including JITing across iterations (e.g., iteration 1 executes 9.5\% more instructions than iteration 5).
The system is based on a standard AOSP 10.0 distribution which does not use transparent huge pages.
We use virtualization, as future mobile systems are expected to use pKVM for increased security~\cite{pkvm}.

\begin{table}[tb]
\centering
\footnotesize
\caption{Mobile-core simulation configuration.}
\label{tab:mobconf}
\begin{tabular}{@{}ll@{}}
\toprule
Processor               & 3GHz Out-of-order Armv8 Processor                        \\
Caches                  & L1 I/D: 32KB 4w, L2: 512KB 8w, L3: 2MB 16w               \\
Memory                  & 90ns, 48GB/s                                             \\ \midrule
L1 TLB                  & I: 32-entry FA, D: 48-entry FA                          \\ 
L2 TLB/PWC              &
  \begin{tabular}[c]{@{}l@{}}
    4KB, full translations: 1536-entry 6w                                          \\
    1GB, 2MB, partial \& full translations: 256-entry 4w
  \end{tabular}\\ \bottomrule
\end{tabular}
\end{table}

\begin{figure}[t]
\includegraphics[width=\columnwidth]{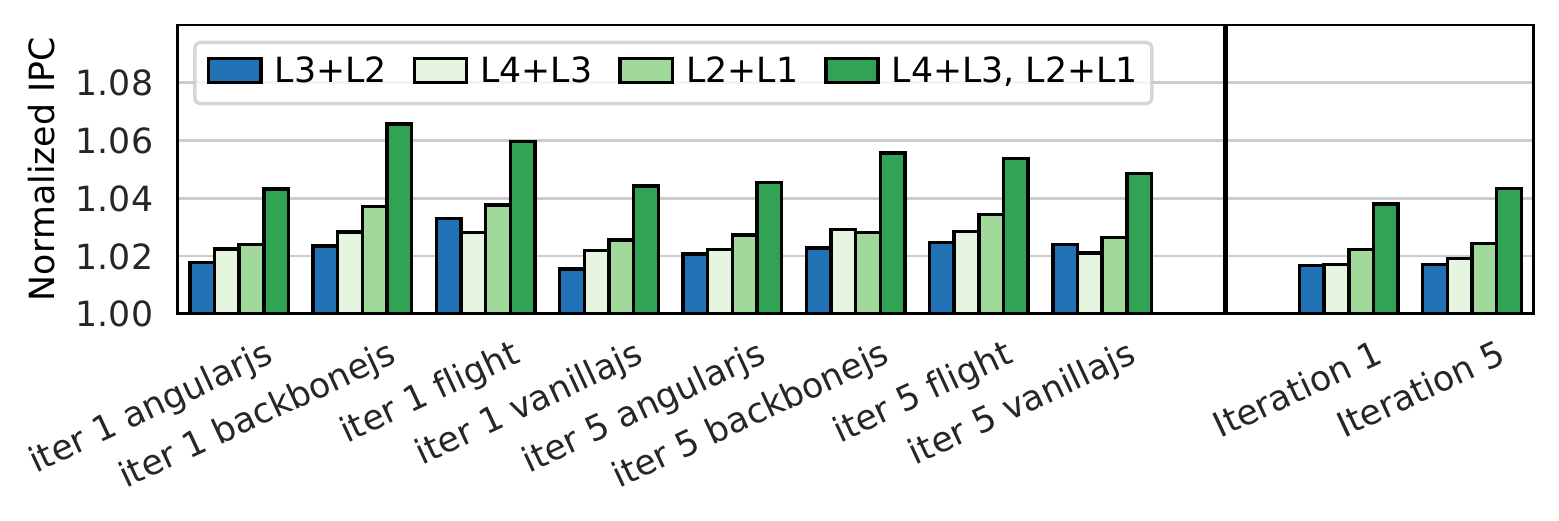}
\caption{Performance of Speedometer 2.0 in a virtualized system normalized to baseline 2-D page table.}
\label{fig:mobA}
\end{figure}

\autoref{fig:mobA} shows the performance gains for a range of flattening options.
The improvement is largest for flattening both L4+L3 and L2+L1
(dark blue bars, 3.8\% and 4.3\% for iterations 1 and 5), which is consistent with our earlier server results.
Overall, flattening closer to the leaf nodes delivers the largest benefit, as they make up the majority of the nodes and are least likely to be cached, particularly under virtualization.

\subsection{Flattening other levels}
We have also simulated flattening the L3+L2 layer, which is by design beneficial for 2\MB data mappings as discussed
in~\autoref{sec:flat:lp}.
For L3+L2 flattening, our results show a 0.2, 0.3, 0.1 percentage point benefit over the
baseline for 0\%, 50\% and 100\% large page scenarios, respectively.
The improvements for virtualization (flattening both host and guest) is 0.7, 1.0, and
1.2 percentage points for the 0\%, 50\% and 100\% large page scenarios.
Finally, for the 100\% large page scenario, we found that L2+L3 flattening outperformed L4+L3 and L2+L1 flattening by 0.3 and 0.8 percentage points, for native and virtualized executions.
These results are consistent with what we saw for the mobile system (\autoref{fig:mobA}).


\section{Conclusion}
In this work we explored two complementary techniques for reducing the impact of page walks: reducing the \textit{number of accesses} by flattening the page table and \textit{reducing the latency} of the accesses by preferentially caching page table entries.
We evaluated server and mobile systems across a wide range of benchmarks with both academic and industrial simulators.

We show that modern PWCs result in little impact from flattening for non-fragmented and non-virtualized systems, but that with the increased page table sizes of realistically fragmented systems and complexity of virtualized page walks, flattening provides significant benefit. Further, we see that preferentially caching page table entries during periods of high TLB miss rates provides significant benefit in all scenarios, as high TLB miss rates are strongly correlated with high data cache miss rates, and the page table is sufficiently smaller than the data that it is far more likely to see reuse through the cache hierarchy. 
We further identify the challenges of self-referencing page tables and provide a practical solution.

Combined, flattening and prioritization allow us to serve the vast majority of page walks with a single cache hit, delivering significant performance (+14.0\%, +7.2\% with realistic large page fragmentation) and dynamic energy (-8.7\% cache and -4.7\% DRAM) benefits beating the state-of-the-art.
Implementation requires only very small changes to the operating system (as we leverage existing
large page support and provide a graceful fallback path) and hardware (as we use existing performance counters and cache partitioning techniques, or need one bit per tag).
If main memory growth continues to outpace TLB growth, we expect that these techniques will become increasingly important.

\section*{Acknowledgement}
This work was supported by the Knut and Alice Wallenberg Foundation through the Wallenberg Academy Fellows Program (grant No 2015.0153), the European Research Council (ERC) under the European Union’s Horizon 2020 research and innovation program (grant No 715283), and the NRF of Korea through the Postdoctoral Fellowship Program (NRF-2020R1A6A3A03037317).

\bibliography{paper.bbl}

\end{document}